\documentclass{siamltex}
\usepackage{balance}
\usepackage[sort,compress]{cite}
\usepackage{graphics,epsfig,graphicx,float,subfigure,color}
\usepackage[section]{algorithm}
\usepackage{amsmath,amssymb,amsbsy,amsfonts}
\usepackage{multicol,multirow}
\usepackage{hhline}
\usepackage{comment}
\usepackage{paralist}
\usepackage{array}
\usepackage{marvosym}
\usepackage{url}
\usepackage{boxedminipage}
\usepackage{threeparttable}
\usepackage{pdfpages}
\usepackage{xspace}
\usepackage{xcolor}

\usepackage{tikz} 
\usetikzlibrary{arrows,decorations.markings,snakes} 
\usetikzlibrary{shapes,calc,shadows,chains}

\newcommand{\p} {\partial}

\definecolor{utorange}{RGB}{203,96,21}
\definecolor{utblack}{RGB}{99,102,106}
\definecolor{utbrown}{RGB}{110,98,89}
\definecolor{utsecbrown}{RGB}{217,200,158}
\definecolor{utsecgreen}{RGB}{208,222,187}
\definecolor{utsecblue}{RGB}{127,169,174}

\tikzstyle{line} = [draw, -latex']

\tikzstyle{vecArrow} = [thick, decoration={markings,mark=at position
   1 with {\arrow[semithick]{open triangle 60}}},
   double distance=2.4pt, shorten >= 5.5pt,
   preaction = {decorate},
   postaction = {draw,line width=2.4pt, white,shorten >= 5.5pt}]
\tikzstyle{innerWhite} = [semithick, white,line width=1.4pt, shorten >= 4.5pt]

\def\abox[#1,#2,#3,#4,#5]#6{%
  \node[draw, cylinder, alias=cyl, shape border rotate=90, aspect=#3, %
  minimum height=#1, minimum width=#2, outer sep=-0.5\pgflinewidth, %
  color=black, left color=olive!30, right color=olive!40, middle
  color=white] (#4) at #5 {};%
  \node at #5 {#6};%
  \fill [olive!20] let \p1 = ($(cyl.before top)!0.5!(cyl.after top)$), \p2 =
  (cyl.top), \p3 = (cyl.before top), \n1={veclen(\x3-\x1,\y3-\y1)},
  \n2={veclen(\x2-\x1,\y2-\y1)} in (\p1) ellipse (\n1 and \n2); }

\def\norm#1{\|#1\|}

\pgfmathsetmacro{\rvec}{.8}
\pgfmathsetmacro{\thetavec}{30}
\pgfmathsetmacro{\phivec}{60} 

\DeclareMathOperator{\sym}{sym}
\DeclareMathOperator{\trace}{tr} 
  
\begin{document}
%

\title{A Nested Partitioning Scheme for Parallel Heterogenous Clusters}

\author{Jesse Kelly\thanks{Institute for Computational Engineering and Sciences, University of Texas, Austin} \and Omar Ghattas \and Hari Sundar}
%

\newcommand{\TODO}[1]{ \fbox{\parbox{3in}{\bf TODO: #1}}}

\newcommand{\grbf}[1] {\mbox{\boldmath${#1}$\unboldmath}}
\newcommand{\gbf}[1] {\mathbf{#1}}

\newcommand{\beq} {\begin{equation}}
\newcommand{\eeq} {\end{equation}}
\newcommand{\bdm} {\begin{displaymath}}
\newcommand{\edm} {\end{displaymath}}
\newcommand{\bit}{\begin{itemize}}
\newcommand{\eit}{\end{itemize}}
\newcommand{\bde}{\begin{description}}
\newcommand{\ede}{\end{description}}
\newcommand{\bce}{\begin{center}}
\newcommand{\ece}{\end{center}}
\newcommand{\ben} {\begin{enumerate}}
\newcommand{\een} {\end{enumerate}}
\newcommand{\bea} {\begin{eqnarray}}
\newcommand{\eea} {\end{eqnarray}}
\newcommand{\barr} {\begin{array}}
\newcommand{\earr} {\end{array}}
\newcommand{\bean} {\begin{eqnarray*}}
\newcommand{\eean} {\end{eqnarray*}}
\newcommand{\edoc} {\end{document}}
\newcommand{\On}{\hspace{0.2cm} \mbox{on} \hspace{0.2cm}}
\newcommand{\Or}{\hspace{0.2cm} \mbox{or} \hspace{0.2cm}}
\newcommand{\In}{\hspace{0.2cm} \mbox{in} \hspace{0.2cm}}
\newcommand{\Minimize}{\hspace{0.2cm} \mbox{minimize} \hspace{0.2cm}}
\newcommand{\Maximize}{\hspace{0.2cm} \mbox{maximize} \hspace{0.2cm}}
\newcommand{\SubjectTo}{\hspace{0.2cm} \mbox{subject} \hspace{0.15cm}
            \mbox{to} \hspace{0.2cm}}

\def\etal{{\it et al.~}}

\newcommand{\point}[1] {\ensuremath{\boldsymbol{#1}}}
\newcommand{\vvec}[1] {\ensuremath{\boldsymbol{#1}}}
\renewcommand{\vec}[1] {\ensuremath{\boldsymbol{#1}}}
\newcommand{\cvec}[1] {\ensuremath{\boldsymbol{{#1}}}}
\newcommand{\ten}[1] {\ensuremath{\boldsymbol{#1}}}
\newcommand{\tenfour}[1] {\ensuremath{\boldsymbol{\mathsf{#1}}}}
\newcommand{\comp}[1]{\begin{pmatrix}{ #1 }\end{pmatrix}}
\newcommand{\vv}{\ensuremath{\vec{v}}}
\newcommand{\vr}{\left(\ensuremath{\vec{r}}\right)}
\newcommand{\att}{\left(t\right)}
\newcommand{\tvv}{\ensuremath{\tilde{\vec{v}}}}
\newcommand{\ww}{\ensuremath{\vec{w}}}
\newcommand{\GG}{\ensuremath{\ten{G}}}
\newcommand{\HH}{\ensuremath{\ten{H}}}
\newcommand{\EE}{\ensuremath{\vec{E}}}
\newcommand{\FF}{\ensuremath{\boldsymbol{\mathfrak{F}}}}
\newcommand{\tEE}{\ensuremath{\tilde{\vec{E}}}}
\newcommand{\nn}{\ensuremath{\vec{n}}}
\renewcommand{\SS}{\ensuremath{\ten{S}}}
\newcommand{\tSS}{\ensuremath{\tilde{\ten{S}}}}
\newcommand{\cp}{\ensuremath{{c_p}}}
\newcommand{\cs}{\ensuremath{{c_s}}}
\newcommand{\tcp}{\ensuremath{{\tilde{c}_p}}}
\newcommand{\tcs}{\ensuremath{{\tilde{c}_s}}}
\newcommand{\Vp}{\ensuremath{{V^e_p}}}
\newcommand{\Vs}{\ensuremath{{V^e_s}}}
\newcommand{\Sp}{\ensuremath{{S^e_p}}}
\newcommand{\Ss}{\ensuremath{{S^e_s}}}
\newcommand{\nxnx}[1] {\ensuremath{\nn\times\left(\nn\times{#1}\right)}}

\newcommand{\dd}[2] {\ensuremath{\frac{\partial {#1}}{\partial {#2}}}}
\newcommand{\DD}[2] {\ensuremath{\frac{d {#1}}{d {#2}}}}
\newcommand{\Grad} {\ensuremath{\nabla}}
\newcommand{\Gradv}[1]{\ensuremath{\nabla_{#1}}}
\newcommand{\Div} {\ensuremath{\nabla\cdot}}
\newcommand{\Divv}[1] {\ensuremath{\nabla_{#1}\cdot}}
\newcommand{\Curl} {\ensuremath{\nabla\times}}
\newcommand{\CC}{\ensuremath{\tenfour{C}}}

\newcommand{\eval}[2][\right]{\relax
  \ifx#1\right\relax \left.\fi#2#1\rvert}

\providecommand{\abs}[1]{\left\lvert#1\right\rvert}
\providecommand{\norm}[1]{\left\lVert#1\right\rVert}

\newcommand{\Nel} {\ensuremath{{n_\text{el}}}}
\newcommand{\De} {\ensuremath{{\mathsf{D}^e}}}
\newcommand{\Dep} {\ensuremath{{\mathsf{D}^{e'}}}}
\newcommand{\Dhat} {\ensuremath{\hat{\mathsf{D}}}}
\newcommand{\Dehat} {\ensuremath{{\hat{\mathsf{D}}^e}}}
\newcommand{\half} {\ensuremath{\frac{1}{2}}}
\newcommand{\IN} {\ensuremath{{\mathcal{I}_N}}}
\newcommand{\INe} {\ensuremath{{\mathcal{I}_{N_e}}}}

\newcommand{\cD}{\mathcal{D}}
\newcommand{\cJ}{\mathcal{J}}
\newcommand{\cL}{\mathcal{L}}
\newcommand{\sR}{\mathbb{R}}
\newcommand{\at}{\tilde{a}}
\newcommand{\ft}{\tilde{f}}
\newcommand{\At}{\tilde{A}}
\newcommand{\pb}{\bar{p}}
\newcommand{\pt}{\tilde{p}}
\newcommand{\ph}{\hat{p}}
\newcommand{\qb}{\bar{q}}
\newcommand{\qh}{\hat{q}}
\newcommand{\sh}{\hat{s}}
\newcommand{\st}{\tilde{s}}
\newcommand{\ub}{\bar{u}}
\newcommand{\uh}{\hat{u}}
\newcommand{\angles}[1]{\left\langle #1 \right\rangle}
\newcommand{\GN}{\mathrm{GN}}
\newcommand{\FN}{\mathrm{FN}}
\newcommand{\SO}{\mathrm{SO}}
\newcommand{\pc} {\cvec{p}}
\newcommand{\gc} {\cvec{g}}
\newcommand{\qc} {\cvec{q}} 

\newcommand{\mc}[1]{\mathcal{#1}}
\newcommand{\mcC}[1]{\mathcal{#1}}
\newcommand{\mb}[1]{\mathbf{#1}}
\newcommand{\mbb}[1]{\mathbb{#1}}
\newcommand{\mi}[1]{\mathit{#1}}
\newcommand{\mf}[1]{\mathfrak{#1}}
\newcommand{\Oi}[1]{\int_{\mc{D}} #1 \, d\vec{x}}
\newcommand{\TOi}[1]{\int_{0}^T \int_{\mc{D}} #1 \, d\vec{x}dt}
\newcommand{\Ti}[1]{\int_{0}^T #1 \, dt}
\newcommand{\TDei}[1]{\int_{0}^T\int_{\De} #1 \, d\vec{x}dt}
\newcommand{\TDeid}[1]{\int_{0}^T\int_{\De,N} #1 \, d\vec{x}dt}
\newcommand{\Dei}[1]{\int_{\De} #1 \, d\vec{x}}
\newcommand{\DeI}[1]{\int_{D} #1 \, d\vec{x}}
\newcommand{\DeiM}[1]{\int_{\Dhat} #1 \, d\vec{r}}
\newcommand{\DeMd}[1]{\int_{\De,N_e} #1 \, d\vec{x}}
\newcommand{\DeiMd}[1]{\int_{\Dhat,N_e} #1 \, d\vec{r}}
\newcommand{\DeMdN}[1]{\int_{D,N} #1 \, d\vec{x}}
\newcommand{\DeiMdN}[1]{\int_{\Dhat,N} #1 \, d\vec{r}}
\newcommand{\DeiMdNC}[2]{\int_{\Dhat,N_{#2}} #1 \, d\vec{r}}
\newcommand{\pDei}[1]{\int_{\partial \De} #1 \, d\vec{s}}
\newcommand{\pDeiM}[1]{\int_{\partial \Dhat} #1 \, d\vec{s}}
\newcommand{\pDeiMd}[1]{\int_{\partial \Dhat,N_e} #1 \, d\vec{s}}
\newcommand{\pDeiMdN}[1]{\int_{\partial \Dhat,N} #1 \, d\vec{s}}
\newcommand{\pDeiMdNC}[2]{\int_{\partial \Dhat,N_{#2}} #1 \, d\vec{s}}
\newcommand{\pMortar}[2]{\int_{\mc{M}_{#2}} #1 \, d\vec{s}}
\newcommand{\pMortard}[2]{\int_{\mc{M}^{#2},N_{#2}} #1 \, d\vec{s}}
\newcommand{\pMortardd}[3]{\int_{\mc{M}^{#2},N_{#3}} #1 \, d\vec{s}}
\newcommand{\pMortardh}[3]{\int_{\mc{M}_{#2},N_{#3}} #1 \, d\vec{s}}
\newcommand{\pMortardhn}[4]{\int_{{#2}_{#3},N_{#4}} #1 \, d\vec{s}}
\newcommand{\TpDei}[1]{\int_{0}^T\int_{\partial \De} #1 \, d\vec{x}}
\newcommand{\TpDeid}[1]{\int_{0}^T\int_{\partial \De,N} #1 \, d\vec{x}}
\newcommand{\Deid}[1]{\int_{\De,N_e} #1 \, d\vec{x}}
\newcommand{\DeidN}[1]{\int_{\De,N} #1 \, d\vec{x}}
\newcommand{\pDeidN}[1]{\int_{\partial \De,N} #1 \, d\vec{s}}
\newcommand{\pDeid}[1]{\int_{\partial \De,N_e} #1 \, d\vec{s}}
\newcommand{\nor}[1]{\left\| #1 \right\|}
\newcommand{\snor}[1]{\left| #1 \right|}
\newcommand{\LRp}[1]{\left( #1 \right)}
\newcommand{\LRs}[1]{\left[ #1 \right]}
\newcommand{\LRa}[1]{\left< #1 \right>}
\newcommand{\LRc}[1]{\left\{ #1 \right\}}
\newcommand{\pp}[2]{\frac{\partial #1}{\partial #2}}
\newcommand{\ppcp}[1]{\frac{\partial #1}{\partial \vec{c_p}}}
\newcommand{\ppcpj}[1]{\frac{\partial #1}{\partial c_p}}
\newcommand{\ppcpmf}[1]{\frac{\partial #1}{\partial \vec{\mf{c_p}}}}
\newcommand{\ppm}[1]{\frac{\partial #1}{\partial \vec{m}}}
\newcommand{\ppmi}[1]{\frac{\partial #1}{\partial m_i}}
\newcommand{\ppmj}[1]{\frac{\partial #1}{\partial m_j}}
\newcommand{\ppcpm}[1]{\frac{\partial #1}{\partial \vec{c_p}^-}}
\newcommand{\ppcpp}[1]{\frac{\partial #1}{\partial \vec{c_p}^+}}
\newcommand{\ppcs}[1]{\frac{\partial #1}{\partial \vec{c_s}}}
\newcommand{\ppcsm}[1]{\frac{\partial #1}{\partial \vec{c_s}^-}}
\newcommand{\ppcsp}[1]{\frac{\partial #1}{\partial \vec{c_s}^+}}
\newcommand{\td}[2]{\frac{d #1}{d #2}}
\newcommand{\Rvert}[1]{\left. #1 \right|}
\newcommand{\bs}[1]{\boldsymbol{#1}}
\newcommand{\ipDed}[3]{\LRp{ #1, #2}_{\De,N}^{#3}}

\newcommand{\jump}[1] {\ensuremath{\LRs{\![#1]\!}}}
\newcommand{\diff}[1] {\ensuremath{\LRs{#1}}}
\newcommand{\average}[1] {\ensuremath{\LRc{\!\{#1\}\!}}}

\newtheorem{propo}{proposition}
\newtheorem{coro}{porollary}
\newtheorem{theo}{theorem}
\newtheorem{lem}{lemma}

\newtheorem{defi}{definition}

\newtheorem{rema}{remark}

\newcommand{\figlab}[1]{\label{fig:#1}}
\newcommand{\eqnlab}[1]{\label{eq:#1}}
\newcommand{\theolab}[1]{\label{theo:#1}}
\newcommand{\corolab}[1]{\label{coro:#1}}
\newcommand{\propolab}[1]{\label{propo:#1}}
\newcommand{\lemlab}[1]{\label{lem:#1}}
\newcommand{\defilab}[1]{\label{defi:#1}}
\newcommand{\remalab}[1]{\label{rema:#1}}

\newcommand{\figref}[1]{\ref{fig:#1}}
\newcommand{\theoref}[1]{\ref{theo:#1}}
\newcommand{\defiref}[1]{\ref{defi:#1}}
\newcommand{\remaref}[1]{\ref{rema:#1}}
\newcommand{\cororef}[1]{\ref{coro:#1}}
\newcommand{\proporef}[1]{\ref{propo:#1}}
\newcommand{\lemref}[1]{\ref{lem:#1}}
\newcommand{\eqnref}[1]{\eqref{eq:#1}}
\newcommand{\alglab}[1]{\label{alg:#1}}
\newcommand{\seclab}[1]{\label{sec:#1}}
\newcommand{\secref}[1]{\ref{sec:#1}}

\maketitle
\begin{abstract}
Modern supercomputers are increasingly requiring the presence of accelerators and co-processors. However, it has not been easy to achieve good performance on such heterogeneous clusters. The key challenge has been to ensure good load balance and that neither the CPU nor the accelerator is left idle. Traditional approaches have offloaded entire computations to the accelerator, resulting in an idle CPU, or have opted for task-level parallelism requiring large data transfers between the CPU and the accelerator. True work-parallelism has been hard as the Accelerators cannot directly communicate with other CPUs (besides the host) and Accelerators. In this work, we present a new nested partition scheme to overcome this problem. By partitioning the work assignment on a given node asymmetrically into boundary and interior work, and assigning the interior to the accelerator, we are able to achieve excellent efficiency while ensure proper utilization of both the CPU and Accelerator resources. The problem used for evaluating the new partition is an $hp$ discontinuous Galerkin spectral element method for a coupled elastic--acoustic wave propagation problem. 
\end{abstract}


\section{Introduction}
\label{sec:intro}

Recent trends in high performance computing points to an ever increasing number of heterogeneous clusters. The trend has been the shift from single core nodes to multi core nodes, and more recently to multi-core nodes with additional cores available via accelerators such as GPUs and the Intel Xeon Phi. Such machines, although capable of achieving a very high throughput on benchmark codes, have been unable to deliver similar levels of performance and scalability for a large class of problems. One of the key reasons for this has been the asymmetry within the cluster, arising from different kinds of execution units (CPU cores vs Accelerator cores), as well as different communication latency and bandwidth. Most work division strategies still work under an assumption of symmetry, leading to suboptimal performance and scalability. The most common (and successful) approach has been to divide work on a per-node basis, requiring one MPI task per node, and dividing the per-node work using ideas from shared-memory parallel programming. This amounts to using OpenMP or pThreads for achieving parallelism on the CPU cores and offloading computationally expensive tasks onto the accelerator. A significant portion of peak-performance on modern clusters is due to the CPUs; for example, for Stampede at the Texas Advanced Computing Center, 2+PFlops out of a total 10PFlops are delivered by the CPUs. 

Realizing good performance on heterogeneous clusters requires good load balance and to ensure that neither the CPU nor the accelerator is left idle. Traditional approaches have offloaded entire computations to the accelerator, resulting in an idle CPU, or have opted for task-level parallelism requiring large data transfers between the CPU and the accelerator. True work-parallelism has been hard as the Accelerators cannot directly communicate with other CPUs (besides the host) and other Accelerators. For pde-based applications, this effectively only leaves the option of using task-based parallelism, resulting in high data exchange between the CPU and the accelerator. Such CPU-Accelerator communication is typically $\mathcal{O}(N)$, where $N$ is the problem size per node. However, inter-node communication (using MPI) is such applications is typically $\mathcal{O}(N^{d-1/d})$, i.e., $\mathcal{O}(N^2/3)$ for 3D problems.
In this work, we present a new nested partition scheme to overcome this problem. By partitioning the work assignment on a given node asymmetrically into boundary and interior work, and assigning the interior to the accelerator, we are able to achieve excellent efficiency while ensure proper utilization of both the CPU and Accelerator resources. The problem used for evaluating the new partition is an $hp$ discontinuous Galerkin spectral element method for a coupled elastic--acoustic wave propagation problem.   

The rest of the paper is organized as follows:
In the following \S\ref{sec:conser} we present a brief overview of linear hyperbolic conservation laws, our problem of interest. In \S\ref{sec:DGSEM}, we describe the discontinuous Galerkin spectral element method. In \S\ref{sec:profile} we summarize results from an initial profiling of our code that guided our code optimization process. In \S\ref{sec:algo} we summarize the new partitioning algorithm and software design of our code. In \S\ref{sec:results}, we present scalability and robustness tests for the proposed partitioning scheme. 


\section{General setting for linear hyperbolic conservation laws}
\label{sec:conser}

We are interested in linear wave equations governed by linear
hyperbolic conservation laws. In the strong form, a general
equation is given as
\begin{equation}
\mc{Q}\pp{{\mf{q}}}{t} + \Divv{\vec{x}} \LRp{\mf{F}{\mf{q}}} = \mf{g} ,
\quad \mf{q} \in \mc{V}, \vec{x} \in \mc{D}, \nonumber
\end{equation}
with $\mc{V}$ as the solution space, to be specified later, over the domain of 
interest $\mc{D}$, and with appropriate initial and boundary conditions. The
subscript $\vec{x}$ denotes the $\vec{x}$-coordinate
system in which the divergence operator acts. Next, multiplying by the test
function $\mf{p}$, the
corresponding weak formulation is obtained as
\begin{align}
\Oi{\mc{Q}\pp{{\mf{q}}}{t}    \cdot {\mf{p}}} + 
\Oi{\Divv{\vec{x}} \LRp{\mf{F}{\mf{q}}} \cdot {\mf{p}}} = 
\Oi{\mf{g}                   \cdot {\mf{p}}},\nonumber
\end{align}
where ``$\cdot$'' denotes the Euclidean inner product.

We next partition the domain $\mc{D}$ into $\Nel$ non-overlapping
hexahedral elements such that $\mc{D} =
\bigcup_{e=1}^{\Nel} \De$, and integrate the weak formulation by parts twice
to obtain
\begin{align}\eqnlab{linearWeakFlux}
&\sum_{e}\Dei{\mc{Q}^e\pp{{\mf{q}}^e}{t}    \cdot {\mf{p}}^e} + 
\Dei{\Divv{\vec{x}} \LRp{\mf{F}{\mf{q}}^e} \cdot {\mf{p}}^e} \nonumber \\ 
&+
\pDei{\vec{n}\cdot 
\LRs{\LRp{\mf{F}{\mf{q}}^e}^*- {\mf{F}{\mf{q}}^e}} \cdot 
{\mf{p}}^e} 
= 
\sum_e\Dei{{\mf{g}}^e                   \cdot \mf{p}^e},
\end{align}
where a consistent numerical flux
 $\LRp{\mf{F}{\mf{q}}^e}^*$ has been introduced to couple solutions of neighboring
elements, and 
$\LRp{\cdot}^e$ denotes the restriction on the $e$-th element
 of the corresponding
quantity.

Equation \eqnref{linearWeakFlux} is known as the strong form in the 
context of nodal discontinuous Galerkin methods
\cite{HesthavenWarburton08}. For the DGSEM described in this paper, the
strong form (integrating the flux terms by parts twice) 
and the weak form (integrating the flux terms by parts once) are 
equivalent \cite{TengLinChangEtAl08, KoprivaGassner10},
and hence all the results in the paper hold for the weak form as well. 

\section{A discontinuous Galerkin spectral element method}
\label{sec:DGSEM} In this section, we briefly describe an {\em
  hp}-discontinuous Galerkin spectral element method. We assume that
each element $\De$ is mapped to the reference hexahedron $\Dhat =
[-1,1]^3$ by a $C^1$-diffeomorphism $\point{X}^e$, and $\mc{D} \approx
\mc{D}^{N_{el}} = \bigcup_{e=1}^{N_{el}}\De$.  Equation
\eqnref{linearWeakFlux} can be written in terms of $\Dhat$ as
\begin{align}\eqnlab{linearWeakFluxMaster}
&\sum_{e}\DeiM{J^e \mc{Q}^e\pp{{\mf{q}}^e}{t}    \cdot {\mf{p}}^e} + 
\DeiM{\Divv{\vec{r}} \LRp{\tilde{\mf{F}}{\mf{q}}^e} \cdot {\mf{p}}^e} \nonumber \\ 
&+
\pDeiM{\tilde{\vec{n}}\cdot 
\LRs{\LRp{\tilde{\mf{F}}{\mf{q}}^e}^*- {\tilde{\mf{F}}{\mf{q}}^e}} \cdot 
{\mf{p}}^e} 
= 
\sum_e\DeiM{J^e{\mf{g}}^e                   \cdot {\mf{p}^e}},
\end{align}
where $\vec{r} = \LRp{r_1,r_2,r_3} \in \Dhat$ 
represents the reference coordinates and $J^e$ is the Jacobian of the
transformation. 
The outward normal on the boundary 
of the master element $\Dhat$ is denoted by $\tilde{\vec{n}}$, and the
contravariant flux \cite{Kopriva06} is defined as
\beq
\tilde{\mf{F}}^i = J^e \vec{a}^i \cdot \mf{F}, \quad i=1,2,3, 
\nonumber
\eeq
with $\vec{a}^i$ as the contravariant basis vectors.

We now describe the approximation spaces for wave propagation in
coupled elastic--acoustic media using the
strain--velocity formulation. Equation
\eqnref{linearWeakFluxMaster} can be specialized to the 
elastic--acoustic wave 
propagation case by the following definitions,
\[
\mf{q} =
  \begin{pmatrix}
    \ten{E} \\ \vvec{v}
  \end{pmatrix}\in \mc{V}, \quad
\mc{Q} =
  \begin{pmatrix}
  \tenfour{I} & \bf{0} \\
  \bf{0} & \rho \ten{I}
  \end{pmatrix}, \quad
\mf{g}  =
  \begin{pmatrix}
    \bf{0} \\ \vvec{f}
  \end{pmatrix}\in \mc{V}, \\
\]
with $\tenfour{I}$ denoting the fourth-order identity tensor, $\bf{0}$
the zero tensors of appropriate sizes, $\ten{I}$ the second-order
identity tensor, $\ten{E}$ the strain tensor, $\vvec{v}$ the velocity
vector, $\vvec{f}$ the external volumetric forces, and $\rho$ the
density. In order to justify the above notations, we note that the
strain tensor $\ten{E}$ with values in $\mathbb{R}^{3\times 3}$ can be
identified with a vector-valued field $\overline{\ten{E}}$ in
$\mathbb{R}^{9}$ (since finite dimensional spaces with the same dimension
are isomorphic) by setting $\overline{\ten{E}}_{\LRs{ij}} =
\ten{E}_{ij}$ with $\LRs{ij} = 3(j-1) + i$. However, to simplify the
notations we use the same symbol $\ten{E}$ for both tensor-valued and
vector-valued fields, and this should be clear in each
context. Similar identifications for other tensors are
straightforward.  These rigorous identifications make our exposition
succinct to leave space for the analysis, which is main focus of the
paper.

The action of the flux operator $\mf{F}$ on the strain--velocity unknowns
$\mf{q}$ can be shown to be \cite{WilcoxStadlerBursteddeEtAl10}
\begin{align*}
(\mf{F}\mf{q})_i & =
  \begin{pmatrix}
    -\half\left( \vvec{v}\otimes\vvec{e}_i +
    \vvec{e}_i\otimes\vvec{v} \right)  \\
    -\left(\tenfour{C}\ten{E}\right)\vvec{e}_i
  \end{pmatrix}\in \mc{V}, \quad \text{for $i=1,2,3$},
\end{align*}
where $\vvec{e}_i$'s are the canonical basis vectors in
$\mathbb{R}^3$, and the second order tensor $
\vvec{v}\otimes\vvec{e}_i$ is the standard dyadic product of vectors
$\vvec{v}$ and $\vvec{e}_i$.  For isotropic linear elasticity, the
strain tensor $\vec{E}$ and the Cauchy stress tensor $\ten{S}$ are
related by the fourth-order constitutive tensor $\tenfour{C}$,
\[
\ten{S} = \tenfour{C}\ten{E} = \lambda\trace(\ten{E})\ten{I} + 2\mu\ten{E},
\]
where
$\lambda$ and $\mu$ are the two Lam\'e constants
characterizing the isotropic constitutive relationship.
The longitudinal wave speed $c_p$ and shear wave speed $c_s$ are
defined in terms of the Lam\'e constants and density by
\[
  c_p = \sqrt{\frac{\lambda + 2\mu}{\rho}}
  \qquad \text{and} \qquad
  c_s = \sqrt{\frac{\mu}{\rho}},
\]
with $c_s = 0$ in acoustic regions by virtue of  $\mu = 0$. 

As in \cite{WilcoxStadlerBursteddeEtAl10}, we choose the solution
space to be $\mc{V} = \vec{V}_{\sym}^{3\times 3}\oplus \vec{V}^3$,
where $\vec{V}$ denotes a space of sufficiently smooth functions
defined on $\mc{D}$ so that \eqnref{linearWeakFlux} makes sense (see
Theorem \theoref{convergence} for examples of $\vec{V}$); and
$\vec{V}_{\sym}^{3\times 3}$ denotes the space of $3\times 3$
symmetric matrices with elements in $\vec{V}$.  The discontinuous
approximation to $\vec{V}$ is given by
\[
\vec{V}_{d} := \{q_{d} \in L^2(\mc{D}^\Nel) : 
\eval{q_{d}}_{\De} \circ \point{X}^e \in
\mbb{Q}_{N_e}(\Dhat)\},
\]
where $\mbb{Q}_{N_e}$ is the tensor product of one-dimensional
polynomials of degree at most $N_e$ on the reference element. It
should be pointed out that the polynomial orders need not be the same
for all directions. Nevertheless, we use the same order for clarity of
the exposition. The numerical solution $\mf{q}_{d} \in
 \mc{V}_{d} = \vec{V}_{d,\sym}^{3\times 3} \oplus \vec{V}_d^3$ restricted on each
element $\De$ is specified as
\[
\eval{\mf{q}_{d}}_{\De} \circ \point{X}^e \in \mc{V}_{d}^e \equiv
\mbb{Q}_{N_e,\sym}^{3\times 3}\oplus \mbb{Q}_{N_e}^3
, \qquad
\point{X}^e : \Dhat \rightarrow \De,
\]
where 
Before introducing the Riemann flux, let us recall the following
standard DG notation for quantities associated with element interfaces:
\[
\jump{\mf{q}} = \mf{q}^+\cdot \vec{n}^+ + \mf{q}^-\cdot \vec{n}^-, \quad
\diff{\mf{q}} = \mf{q}^- - \mf{q}^+, \quad 
\average{Z} = \frac{Z^+ + Z^-}{2},
\]
where the positive and negative signs indicate element interior and
exterior, respectively. 

For linear conservation laws one can solve the Riemann problem exactly
by various methods \cite{Toro99}. Using the Rankine--Hugoniot approach, Wilcox \etal
 \cite{WilcoxStadlerBursteddeEtAl10} show that the exact Riemann flux
for the strain equation is given by
\begin{align*}
  \vec{n}\cdot \LRs{
\LRp{\mf{F}{\mf{q}}}^*_{\EE}-\LRp{\mf{F}{\mf{q}}}_{\EE}} =&
   \left(k_0\nn\cdot\jump{\CC\EE} + k_0\rho^+c_p^+\jump{\vv}\right)\nn\otimes\nn\\
  & - k_1\sym\left(\nn\otimes\left(\nn\times(\nn\times \jump{\CC\EE})\right)\right)\\
  & - k_1\rho^+c_s^+\sym\left(\nn\otimes\left(\nn\times(\nn\times \diff{\vv})\right)\right),\\[1ex]
\intertext{and for the velocity equation by}
 \vec{n}\cdot \LRs{\LRp{\mf{F}{\mf{q}}}^*_{\vv}-\LRp{\mf{F}{\mf{q}}}_{\vv}}=&
   \left(k_0\nn\cdot\jump{\CC\EE} + k_0\rho^+c_p^+\jump{\vv}\right)\rho^-c_p^-\nn\\
  & - k_1\rho^-c_s^-\nn\times(\nn\times \jump{\CC\EE})\\
  & - k_1\rho^+c_s^+\rho^-c_s^-\nn\times(\nn\times \diff{\vv}),
\end{align*} 
with
$
  k_0 = \LRp{\rho^-c_p^-+\rho^+c_p^+}^{-1},
$
$ k_1 = \LRp{\rho^-c_s^-+\rho^+c_s^+}^{-1}$ if $\mu^{-} \ne 0$, and 
$k_1 = 0$ if $\mu^- = 0$.
Here, we will consider only traction boundary conditions $\ten{S}\nn
= \bs{t}_{bc}$, where $\bs{t}_{bc}$ is the prescribed traction. The traction condition
is enforced by the following mirror principle,
\begin{align*}
\jump{\vv} = \diff{\vv} = {\bf{0}}, \quad \text{ and }\quad
\jump{\ten{S}} = -2\LRp{\bs{t}_{bc} - \ten{S}^-\nn},
\end{align*}
which applies to both elastic and acoustic media. 

In order to unify the treatment for elastic, acoustic, coupled
elastic--acoustic, and electromagnetic waves, we define a generic
polynomial space $\mc{P}_N$ whose meaning will be clear in each
context. For example, if we write $\mf{q}^e \in \mc{P}_N$, this
identifies $\mc{P}_N \equiv \mc{V}_d^e$.

The tensor product basis for $\mbb{Q}_N$ is built upon the following
one-dimensional Lagrange basis
\[
\ell_l(\xi) = \prod_{\substack{k = 0, 1, \ldots, N\\k \ne l}} \frac{\xi -
  \xi_k}{\xi_l - \xi_k},
\]
where the $N$th-degree 
Legendre-Gauss-Lobatto (LGL) points, or $N$th-degree Legendre-Gauss 
points (LG),
 $\{\xi_l\}$ on $[-1,1]$ for
$l=0,\hdots,N$,  are chosen as both
the interpolation and quadrature points. This is also known as 
the collocation approach. 
The Lagrange interpolant of a function $f(\point{r})$ on the
reference element $\Dhat$ is defined through the interpolation
operator $\IN$ as
\[
\IN \LRp{f} = \sum_{l,m,n=0}^{N} f_{lmn}
\ell_l(r_1) \ell_m(r_2) \ell_n(r_3),
\quad
f_{lmn} = f \left( \point{\xi}_{lmn} \right),
\quad
\point{\xi}_{lmn} =
\begin{pmatrix}
  \xi_l, \xi_m, \xi_n
\end{pmatrix} \in \Dhat.
\]

A typical collocation approach \cite{Kopriva09}
 in semi-discretizing \eqnref{linearWeakFluxMaster}  
is as follows. Find $\mf{q} \in \mc{V}_{d}$ such that
\begin{align}
&\sum_{e}\DeiMd{\INe\LRp{\INe\LRp{J^e} \INe\LRp{\mc{Q}^e}\pp{{\mf{q}}^e}{t}} 
\cdot {\mf{p}}^e} + 
\DeiMd{\Divv{\vec{r}} \INe\LRp{\tilde{\mf{F}}{\mf{q}}^e} \cdot {\mf{p}}^e} \nonumber \\ 
&+
\pDeiMd{\tilde{\vec{n}}\cdot 
\LRs{\INe\LRp{\LRp{\tilde{\mf{F}}{\mf{q}}^e}^*}- \INe\LRp{{\tilde{\mf{F}}{\mf{q}}^e}}} \cdot 
{\mf{p}}^e} 
\nonumber \\
&=\sum_e\DeiMd{\INe\LRp{\INe\LRp{J^e}\INe\LRp{\mf{g}^e}}    \cdot {\mf{p}^e}},
\quad \forall \mf{p} \in \mc{V}_{d}, \eqnlab{MasterDiscrete}
\end{align}
where
$
\INe\LRp{\tilde{\mf{F}}^i} = \INe\LRp{\INe\LRp{J^e \vec{a}^i} \cdot \INe\LRp{\mf{F}}}.
$
The direct consequence of the above collocation is that the integrand 
in each integral is at most of order $2N_e$ in each direction $r_i, i=1,2,3$. The subscript $N_e$ in the integrals 
means that the integrals are numerically evaluated
using the corresponding $N_e$th-degree LGL (or LG) quadrature rule. 
corollary


\section{Profiling}
\label{sec:profile}

A baseline analysis of the original, un-modified \textit{dgae} code was performed. \textit{mangll} and \textit{dgae} were compiled using \textit{mpicc}, with the \textbf{-O3}, and \textbf{-g} flags set. This configuration is identical to what an end-user of \textit{mangll} and \textit{dgae} would compile and run. No explicit vectorization was inserted programmatically, however the compiler was free to vectorize and optimize aggressively. OpenMP was not used in the baseline code.

\begin{figure}[htbp]
\begin{center}
\includegraphics[width=4in]{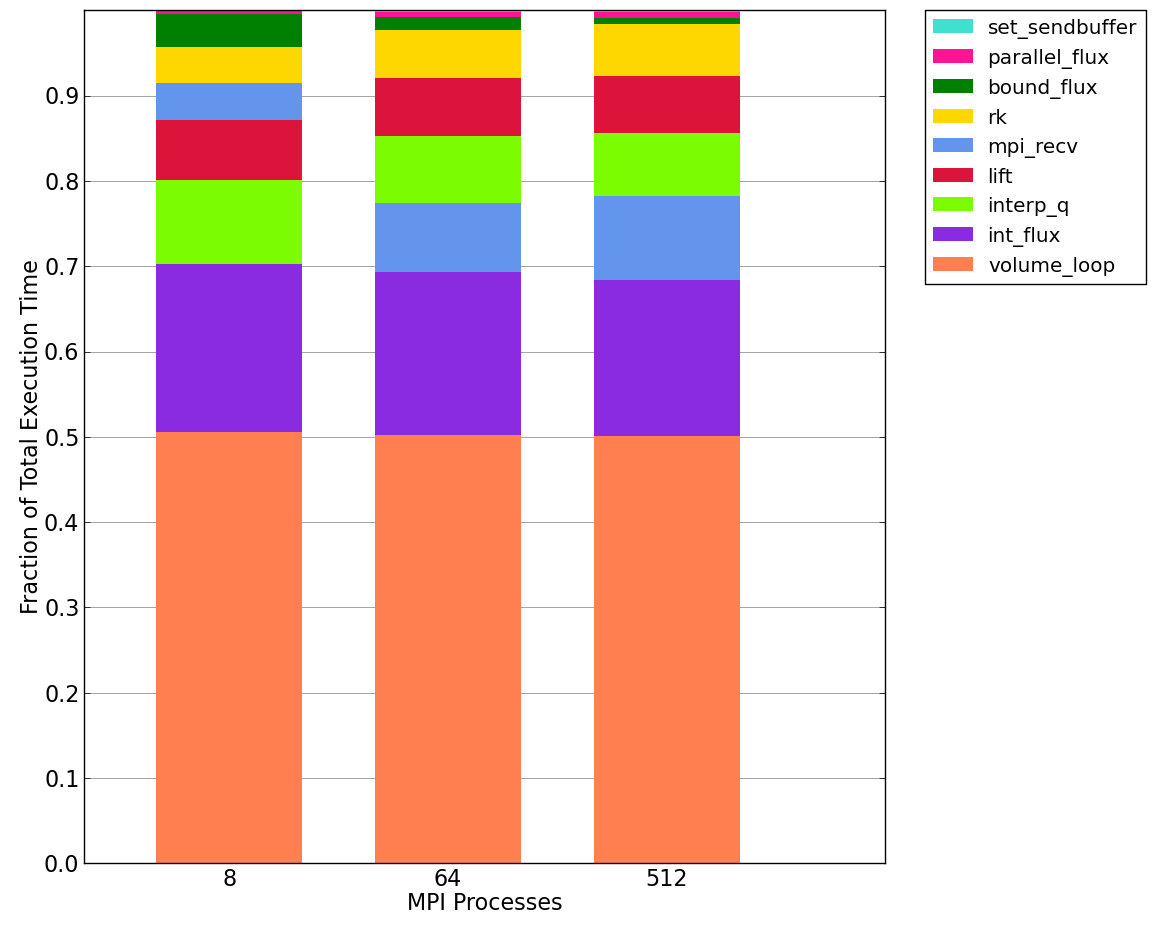}
\caption{Breakdown of total execution time for multiple runs of baseline MPI-only \textit{snell} code. \textbf{Average} refers to the average percentage of total execution time across total number of MPI processes.}
\label{fig:basebreak}
\end{center}
\end{figure}

Using one MPI process per core, a 7th order discretization, and using only one socket on each compute node to control for NUMA effects, the \textit{snell} program was run and profiled on 1, 8, and 64 nodes on Stampede. One socket of a single node contains 8 cores, and so 8 MPI processes per node were used. Each run was executed for 118 timesteps, with 1024 elements per MPI process. These runs revealed several computationally-intensive kernels in the acoustic-elastic wave solution. A breakdown of total program execution time is shown in Figure \ref{fig:basebreak}. 

The \textbf{volume\_loop} kernel encapsulates the elemental tensor product application to each of the nine unknowns. For each unknown, three tensor applications are performed, IIAX, IAIX, and AIIX. Each of these three kernels amounts to \(M\) (= Order + 1) matrix multiplications, each matrix multiplication being one \(M\times M\) matrix times another, where \(M\) is the elemental order plus one. These tensor applications make up the majority of \textbf{volume\_loop}'s, as well as \textit{snell}'s, runtime.

The \textbf{int\_flux} kernel encapsulates the calculation of the fluxes across element faces between elements owned by each MPI process. For each pair of faces in the face mesh, \textbf{int\_flux} calls the \textbf{godonov\_flux} kernel. This kernel is straightforward and embarrassingly parallel within each face, and amounts to various operations performed with vectors of length \(NFP\), where \(NFP = M^3\).

The remaining kernels, \textbf{interp\_q}, which interpolates volume unknowns to the face mesh, \textbf{lift}, which connects the facial fluxes to element interiors, \textbf{rk}, the Runge-Kutta update, and \textbf{bound\_flux} and \textbf{parallel\_flux}, which calculate fluxes on the physical and inter-processor boundaries, respectively, though not as computationally intensive as \textbf{volume\_loop} or \textbf{int\_flux}, still made up a significant enough portion of total execution time to merit investigating the benefits of vectorization. We will see that all kernels require multithreading using OpenMP in order to achieve overall performance comparable to the pure MPI code.

\section{Algorithms \& Design}
\label{sec:algo}

\subsection{Baseline Code}
The original, baseline code written to solve the acoustic-elastic wave equation is called \textit{dgae}, for \textit{discontinuous Galkerin acoustic-elastic} wave solution. This code uses MPI to attain distributed-memory parallelism on traditional supercomputing architectures. No OpenMP threading is used, and no explicit vectorization exists within the computationally intensive kernels that make up the solution of the acoustic-elastic wave equation. A given global geometry is discretized using octahedral finite elements contained in an octree. The finite element discretization of the problem is handled using the \textit{mangll} library \cite{BursteddeGhattasStadlerEtAl09b}. A discretization of a given domain may consist of multiple octrees. The elements can be ordered according to a global Morton ordering \cite{SundarSampathBiros08}, in effect producing a one-dimensional array of elements which is then spliced into roughly equally-sized sub-arrays. Each of these sub-arrays corresponds to a geometric subdomain, which is assigned to an MPI process. This procedure is approximately optimal with respect to minimizing communication between subdomains \cite{SundarSampathBiros08}.

\subsection{Stampede}
The Stampede supercomputer houses 6400 compute nodes, each of which consists of two 8-core Intel Xeon E5 (Sandy Bridge, hereafter SNB) processors and a single 61-core Intel Xeon Phi (Many-Integrated-Core, hereafter MIC) coprocessor. Each Sandy Bridge core runs at 2.7GHz and has 16 256-bit vector registers. Each core can execute one double-precision floating-point vector addition or subtraction and one double-precision floating-point vector multiplication per clock cycle, provided the addition and the multiplication are independent. Hyperthreading is disabled on Stampede, and each SNB core supports one hardware thread. This amounts to a theoretical peak performance of 346 double-precision gigaFLOPs per compute node from the Sandy Bridge cores. The CPU-side memory subsystem on a Stampede compute node consists of four channels per socket, each rated at 1600 MT/s, for a theoretical peak bandwidth of 51.2 GB/s.

Each MIC core runs at 1.1 GHz and has 32 512-bit vector registers per hardware thread, with support for up to four hardware threads. These cores can execute two independent 8-wide double-precision vector floating point operations per clock cycle for a theoretical peak performance of 1.0 double-precision teraFLOPs per compute node. The coprocessor memory subsystem consists of 8 memory controllers serving all the cores iwith a theoretical peak of 320 GB/s. Each coprocessor core has its own private L1 cache, and all cores share a common L2 cache. The cores may communicate directly with each other, bypassing the memory subsystem, through a ring interconnect.

\subsection{Offload Programming}
One option for work sharing on Stampede is to use offloading. In the offload model, a host process on the Sandy Bridge side of the compute node launches kernels on the MIC through offload regions. This programming mode parallels the way a GPU is used in a coprocessing environment. Offload regions can contain perform computation or data-transfer. In this model, the MIC only communicates with the host process that runs on same physical compute node. Though CPU-to-MIC memory transfer still incurs an overhead large enough to necessitate minimizing communication, the overhead is far more manageable than the MIC-to-Infiniband overhead.

\subsection{Mapping to Stampede}
The original \textit{dgae} code was written with the intention of executing the code on traditional supercomputers, with the finest level of parallelism addressed being at the MPI process level. With the increasing size of vector registers and increasing width of vector-processing-units, the performance gap between vectorized and non-vectorized code is becoming wider and wider. The \textit{dgae} and \textit{mangll} codes have traditionally relied on the compiler to provide vectorization for eligible loops. While this certainly provides a performance benifit, we chose to implement hand-coded vectorized kernels using assembly language and vector intrinsics to ensure efficient vectorization. It is especially important to ensure efficient vectorization for the MIC, since there is an eightfold difference in the performance of a vector operation versus a scalar operation. Vectorization of key kernels constitutes the finest level of parallelism considered in producing high-performance code for Stampede.

The second level of parallelism considered is thread-level parallelism. In contrast to the original \textit{dgae} and \textit{mangll} codes, which make use of all available compute cores by assigning one MPI process to each core, the Stampede-optimized versions of these codes institute intra-node, thread-level parallelism using OpenMP, with one OpenMP process running on each core, and institute intra-node-level parallelism using MPI, with one MPI process assigned to each compute node. Each MPI process is responsible for spawning the OpenMP threads on its compute node. The reason for taking this approach is entirely due to attaining higher performance on the MIC.

Each compute node is heterogeneous, with the bulk of the processing power existing on the coprocessor. While using MPI for intra-node communication is feasible on the Sandy Bridge side of the compute nodes, mapping one or more MPI processes to the coprocessor leads to a number of difficulties. The MIC has limited memory, only 8 GB of RAM, compared to the CPU's 32 GB. Using the existing segmentation framework present in \textit{mangll} results in each MPI process owning roughly equally-sized subdomains. If each subdomain was sized such that 61 subdomains could fit in MIC memory, then the CPU would be vastly underutilized. If, instead, each subdomain was guaranteed to provide enough work to the CPU, 61 of such subdomains could not fit in the MIC's memory. Another problem is that processes on the MIC that communicate with off-chip processes via MPI incur a very high communication overhead. 61 such processes all contending for PCI bandwidth would flood each node's PCI-bus, aside from the nightmarish Infiniband traffic generated from 77 MPI processes per compute node all trying to communicate at once. Since overall we seek to minimize inter-processor communication, using smaller subdomains would also increase the communication-to-computation ratio, since in dG it is only subdomain face data that is communicated. 

A more streamlined solution, and one that provides for the usage of \textit{mangll}'s existing homogenous load balancing capability, is to give each compute node a single subdomain, each of which is roughly equal in size. Each compute node's MPI process handles inter-node communication, the spawning of OpenMP threads on the CPU side, and the initiation of offloaded processes. In order to minimize CPU-MIC communication overhead and avoid over-scheduling the MIC, it is best to have only a single process on the CPU-side that launches offload kernels.

\subsection{Partitioning Scheme}
As previously discussed, in the discontinuous Galerkin finite element method, inter-process communication consists entirely of the transfer of shared element faces. Thus, the partitioning of the global domain into processor-local domains directly controls the amount of communication required. If partitions are created such that the number of faces shared amongst processors is minimized, then communication will also be minimized. Partitioning must take place at two levels, then. First, the inter-node partitioning proceeds as has just been described. We assign one MPI process to each compute node, and so each MPI process owns a subdomain of adjacent elements. Next, we must partition each process' domain into elements that will be handled by the CPU and elements that will be offloaded to the MIC. 
 
There are three primary considerations in the generation of the CPU-MIC partitioning. First, we only allow interior elements (that is, elements whose faces are not shared with other compute nodes) to be offloaded to the MIC. Second, we seek to minimizing communication over the PCI bus, which is achieved by minimizing the surface area of the partition offloaded to the MIC. Third, we must consider the heterogeneous nature of load balancing between the CPU and the MIC. We expect the MIC to complete its computation in less time than the CPU for a given number of elements. The more elements the MIC owns, however, the higher the PCI communication overhead becomes. We must strike a balance between the increased compute power of the MIC and the overhead of transferring data across the PCI bus.

This approach differs significantly from the way in which a coprocessor is commonly used. A common paradigm for coprocessor computing is to offload whole computationally intensive tasks to the coprocessor. For exaxmple, in the solution of the acoustic-elastic wave equation, one might offload the \textbf{volume\_loop} to the coprocessor while the CPU computes the \textbf{int\_flux} kernel. While the \textbf{volume\_loop} kernel is certainly computationally intensive, and maps well to coprocessor architectures, the amount of data that must be transferred at each timestep is proportional to the total number of elements $K$ owned by the compute node times the number of degrees of freedom per element $(N+1)^3$. In contrast, transferring only a subset $K_{co}$ of elements to the coprocessor and transferring only shared face data, assuming that $K_{co}$ is structured such that interfacial area is minimized, is proportional to $6K^{2/3}(N+1)^2$. The tradeoff is that, although less data is transferred, we must ensure the entire timestep runs optimally on the MIC, rather than just one or two computationally-intensive kernels. 

\begin{figure}

	\centering
	\includegraphics[width=\textwidth]{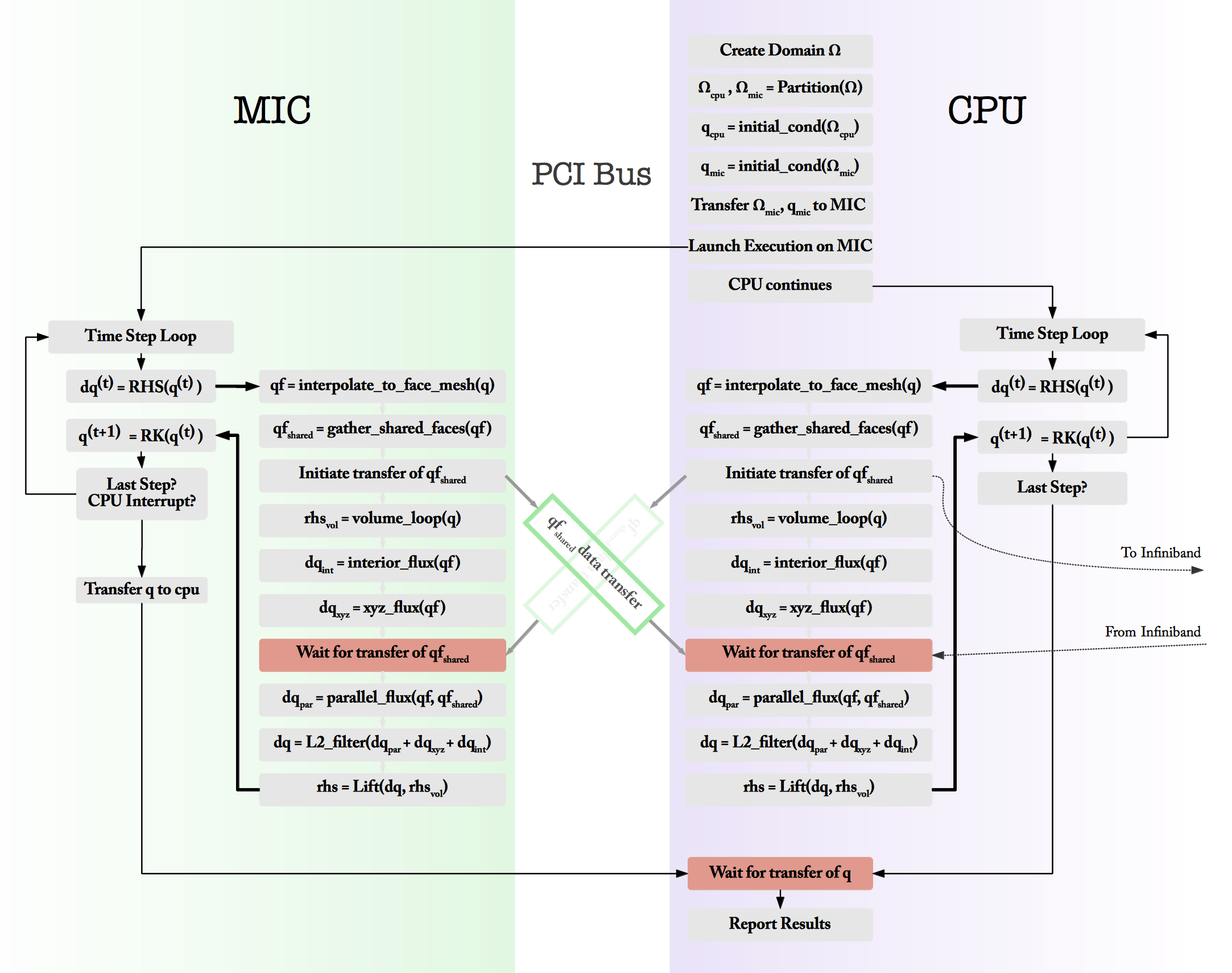}
	\caption{The host-coprocessor execution flow.}
	\label{fig:flowchart}
\end{figure}

The approach proposed in this paper is to treat the dynamic between the host and offload processes in much the same way as the dynamic between compute nodes is treated. Each host initailizes its coprocessor, then launches an offload process that runs each timestep in parallel with the host. Syncronization is only required once per time step, when the CPU and coprocessor exchange their shared face data. At the end of the computation, all of the degrees of freedom computed on the coprocessor are transferred back to the host, and so we incur the cost of transferring large datasets only once. This process is shown in Figure \ref{fig:flowchart}.

\subsection{CPU-MIC Load Balancing}
\label{sec:load_balancing}
\begin{figure}[htbp]
\begin{center}
\includegraphics[width=4in]{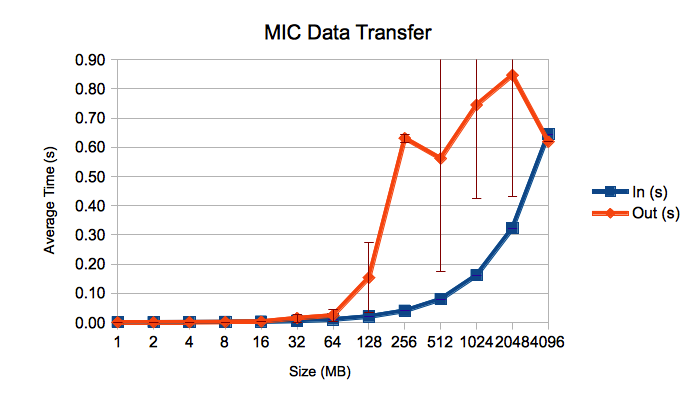}
\caption{Data transfer times between CPU and MIC. Vertical bars show standard deviations in samples.}
\label{fig:mic_xfer}
\end{center}
\end{figure}

To facilitate proper CPU-MIC load balancing, several experiments were run at different orders (N) and numbers of elements (K). For each N, K combination, execution times were recorded for each of the kernels in \textit{dgae} on both the CPU and the MIC. Additionally, memory transfer experiments between the CPU and the MIC were carried out. The time required to transfer an array of double-precision floating point numbers both to and from the MIC was observed for array sizes from 1 to 4096 megabytes (see figure \ref{fig:mic_xfer}). 

\begin{figure}[htbp]
\begin{center}
\includegraphics[width=4in]{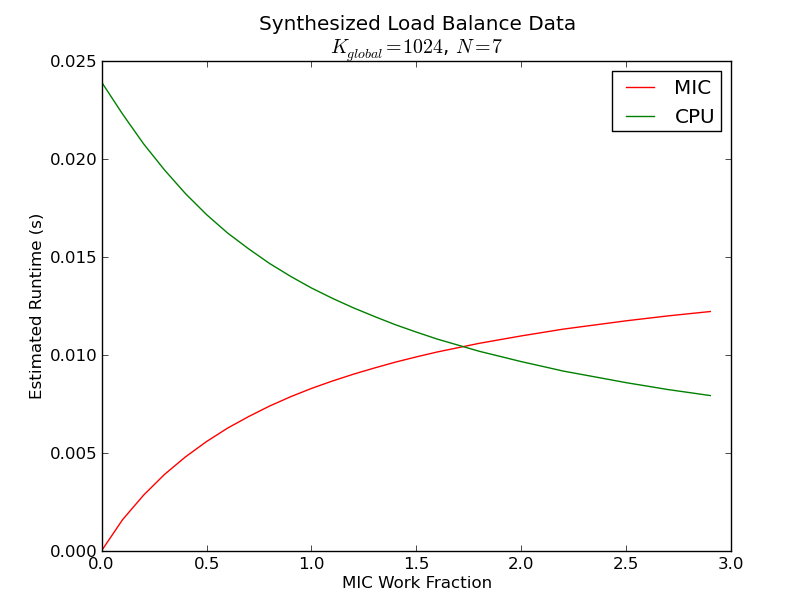}
\caption{Estimated runtimes on CPU and MIC for various load fractions. The point at which the two curves cross shows the optimal MIC work fraction for this N, K combination.}
\label{fig:micbalance}
\end{center}
\end{figure}

This data was used to produce two functions, one for the MIC $T_{MIC}(N, K)$ and one for the CPU $T_{CPU}(N, K)$, that predict the time required to process K N-th order elements for a single time step. $$T_{MIC}(N, K_{MIC}) = \sum kernel_{time,mic}(N, K_{MIC})$$ and $$T_{CPU}(N, K_{CPU}) = \sum\left[kernel_{time,cpu}(N, K_{CPU})\right] + PCI_{time}(K_{MIC})$$
$PCI_{time}(K_{MIC})$ returns an estimate of the time required to transfer shared faces between the CPU and the MIC if there are $K_{MIC}$ elements on the MIC if the surface area of the MIC partition is at a minimum.

Computation on the MIC takes place asynchronously with respect to the CPU host process. Therefore, our load balance will be optimal when $T_{MIC}=T_{CPU}$. For each N, K combination, then, we simply solve the system:

$$T_{MIC}(N, K_{MIC})=T_{CPU}(N, K_{CPU})$$
$$K=K_{MIC}+K_{CPU}$$

where $K$ is the total number of elements in a given compute-node's partition.

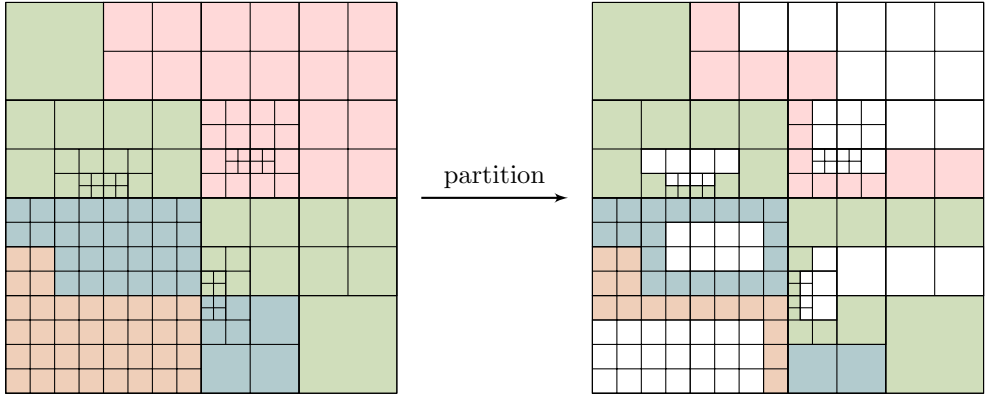
\begin{figure}
	
	\begin{center}
	  \begin{tikzpicture}[scale=0.65]
    
	    \draw[fill=utorange!30] (0,0) rectangle +(4,2);
	    \draw[fill=utorange!30] (0,2) rectangle +(1,1);
		
	    \draw[fill=utsecblue!60] (0,3) rectangle +(1,1);
	    \draw[fill=utsecblue!60] (1,2) rectangle +(3,2);
	    \draw[fill=utsecblue!60] (4,0) rectangle +(2,2);
    
	    \draw[fill=utsecgreen] (6,0) rectangle +(2,2);
	    \draw[fill=utsecgreen] (4,2) rectangle +(4,2);
	    \draw[fill=utsecgreen] (0,4) rectangle +(4,2);
	    \draw[fill=utsecgreen] (0,6) rectangle +(2,2);

	    \draw[fill=red!15] (4,4) rectangle +(4,4);
	    \draw[fill=red!15] (2,6) rectangle +(2,2);
    
	    \draw[step=2] (0,0) grid +(8,8);

	    \draw[step=1] (4,0) grid +(2,4);
	    \draw[step=1] (0,4) grid +(4,2);
	    \draw[step=1] (4,4) grid +(4,4);
    
	    \draw[step=1] (6,2) grid +(2,2);
	    \draw[step=1] (2,6) grid +(2,2);
    
	    \draw[step=0.5] (0,0) grid +(4,4);
    
	    \draw[step=0.5] (4,4) grid +(2,2);
	    \draw[step=0.5] (1,4) grid +(2,1);
	    \draw[step=0.5] (4,1) grid +(1,2);

	    \draw[step=0.25] (1.5,4) grid +(1,0.5);
	    \draw[step=0.25] (4,1.5) grid +(0.5,1);
	    \draw[step=0.25] (4.5,4.5) grid +(1,0.5);
		
			\draw[-latex',thick] (8.5,4) -- node[above] {partition} (11.5,4);
		
			\begin{scope}[shift={(12,0)}]
		    \draw[fill=utorange!30] (0,0) rectangle +(4,2);
		    \draw[fill=utorange!30] (0,2) rectangle +(1,1);
    
				\draw[fill=white] (0,0) rectangle +(3.5,1.5);
		
		    \draw[fill=utsecblue!60] (0,3) rectangle +(1,1);
		    \draw[fill=utsecblue!60] (1,2) rectangle +(3,2);
		    \draw[fill=utsecblue!60] (4,0) rectangle +(2,1);
    
				\draw[fill=white] (1.5,2.5) rectangle +(2,1);
		
		    \draw[fill=utsecgreen] (6,0) rectangle +(2,2);
		    \draw[fill=utsecgreen] (4,2) rectangle +(4,2);
				\draw[fill=utsecgreen] (4,1) rectangle +(2,1);
		    \draw[fill=utsecgreen] (0,4) rectangle +(4,2);
		    \draw[fill=utsecgreen] (0,6) rectangle +(2,2);

				\draw[fill=white] (4.5,2.5) rectangle +(0.5,0.5);
				\draw[fill=white] (4.25,1.5) rectangle +(0.75,1);
				\draw[fill=white] (5,2) rectangle +(3,1);
				\draw[fill=white] (1,4.5) rectangle +(2,0.5);
				\draw[fill=white] (1.5,4.25) rectangle +(1,0.25);
			
		    \draw[fill=red!15] (4,4) rectangle +(4,4);
		    \draw[fill=red!15] (2,6) rectangle +(2,2);
    	
				\draw[fill=white] (5,6) rectangle +(3,2);
				\draw[fill=white] (3,7) rectangle +(2,1);
				\draw[fill=white] (4.5,5) rectangle +(3.5,1);
				\draw[fill=white] (4.5,4.5) rectangle +(1.5,0.5);			
					
		    \draw[step=2] (0,0) grid +(8,8);

		    \draw[step=1] (4,0) grid +(2,4);
		    \draw[step=1] (0,4) grid +(4,2);
		    \draw[step=1] (4,4) grid +(4,4);
    
		    \draw[step=1] (6,2) grid +(2,2);
		    \draw[step=1] (2,6) grid +(2,2);
    
		    \draw[step=0.5] (0,0) grid +(4,4);
    
		    \draw[step=0.5] (4,4) grid +(2,2);
		    \draw[step=0.5] (1,4) grid +(2,1);
		    \draw[step=0.5] (4,1) grid +(1,2);

		    \draw[step=0.25] (1.5,4) grid +(1,0.5);
		    \draw[step=0.25] (4,1.5) grid +(0.5,1);
		    \draw[step=0.25] (4.5,4.5) grid +(1,0.5);
			\end{scope}
			
		\end{tikzpicture}
	\end{center}

	\caption{New partitioning scheme. Each compute node is assigned a portion of the global domain. Subdomains owned by the MIC are then generated within each process' domain by selecting an appropriate number of interior elements to ensure proper load balance. White areas in the second mesh above show subdomains offloaded to each process' MIC.}
	
\end{figure}

\section{Results}
\label{sec:results}

\begin{figure}
\begin{center}
\includegraphics[width=4in]{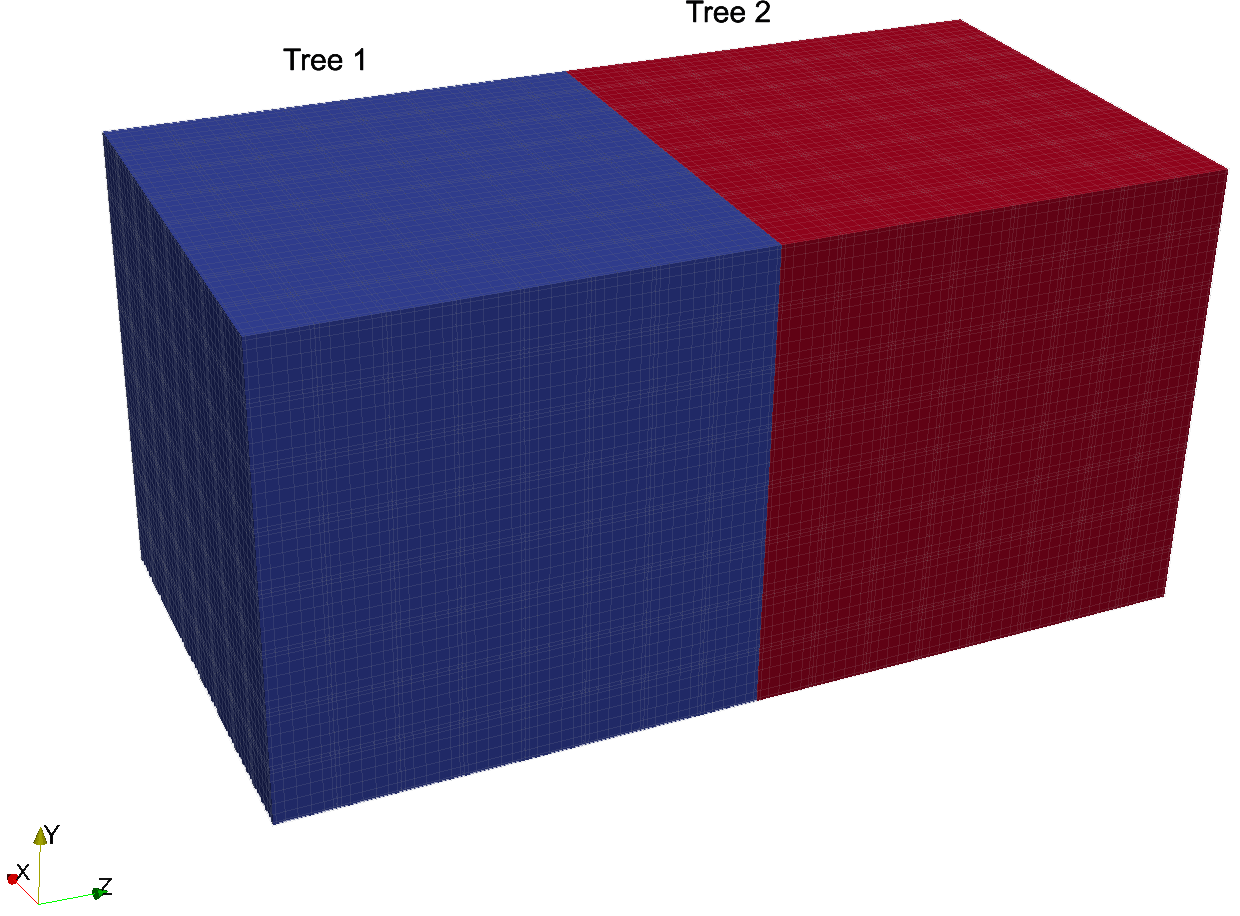}
\caption{Geometry on which the acoustic-elastic wave equation was solved. The first tree had material parameters $c_p=1$ and $c_s=0$, while the second tree had material parameters $c_p=3$ and $c_s=2$.}
\label{fig:snell_setup}
\end{center}
\end{figure}

Table \ref{tab:runtime_comparison} below compares the total execution times for two solutions of the acoustic-elastic wave equation over a brick-like domain with a discontinuity in material parameters at the center of the brick \ref{fig:snell_setup}. Each run was executed with 8192 elements per compute-node. For the basline, pure-MPI run, 8 MPI processes were used per node. For the optimized (vectorized, OpenMP threaded, MIC offloading) runs, 1 MPI process per compute node was used. This process then spawned 8 OpenMP threads per node, each pinned to a separate core on a single socket, and controlled offloading to the MIC. Each offloaded process used 120 threads on the MIC, or two threads per MIC core. The MIC work fraction was calculated using the procedure discussed in Section \ref{sec:load_balancing}, and was found to be $K_{MIC}/K_{CPU} = 1.6$.

\begin{table}[htdp]
\caption{Runtime comparison for original MPI-only code and MIC-enabled, vectorized, hybrid MPI-OMP code. Baseline runs completed with 8 MPI processes per node, optimized runs completed with 1 MPI process, 8 OpenMP threads, and 120 MIC threads per node.}
\begin{center}
\begin{tabular}{|c|c|c|c|}
\hline
Nodes & Baseline Wall Time (s) & Optimized Wall Time (s) & Speedup \\
\hline
   1  & 408                & 65                          & 6.3x    \\
   64 & 413                & 74                          & 5.6x    \\
\hline
\end{tabular}
\end{center}
\label{tab:runtime_comparison}
\end{table}

\begin{figure}
\begin{center}
\includegraphics[width=4in]{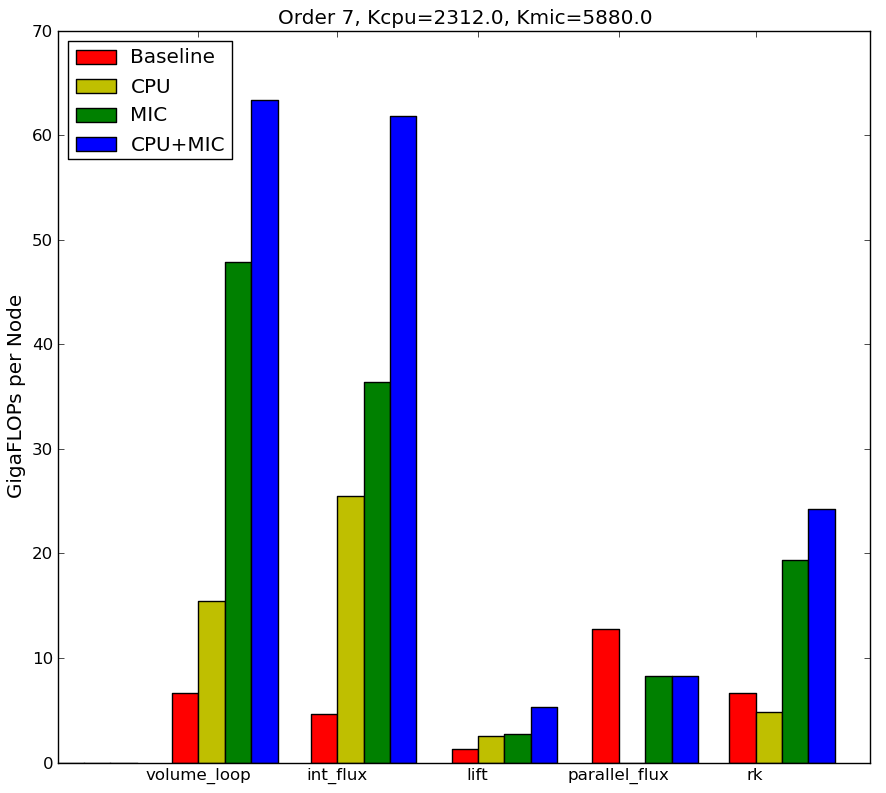}
\caption{Single-node performance comparison. Baseline runs completed with 8 MPI processes. Optimized runs completed with 1 MPI process, 8 OpenMP threads, 120 MIC threads. ``CPU'' refers to the CPU-side execution in the optimized code.}
\label{fig:flop_compare}
\end{center}
\end{figure}

Figure \ref{fig:flop_compare} provides detail concerning baseline versus optimized performance for key kernels. The \textbf{volume\_loop} and \textbf{int\_flux} kernels saw the largest increase in performance between baseline and optimized codes. These kernels, as previously discussed, had huge potential for both vector-level and thread-level parallelism. MIC performance aside, the multithreaded, hand-vectorized versions of these kernels outperform the basline code by a factor of 2x (for \textbf{volume\_loop} and 5x (for \textbf{int\_flux}). Performance on the MIC is consistently higher for all kernels, with the exception of the \textbf{parallel\_flux} kernel. It should be noted that this kernel handles the flux computation on faces shared with other processes. In the baseline case, there was much work to do due to the high number of shared faces between the MPI processes. In the optimized case, the only shared faces were those transmitted across the PCI bus to the MIC.

Each compute node on Stampede has a theoretical peak performance of 173 GFLOPs per socket and 1 TFLOPs per coprocessor. Full utilization of both resources constitute a theoretical peak performance of 1173 GFLOPs per compute node using a single core, or a 6.7x performance benefit from using the MIC compared to simply using a single socket on the CPU side. We have observed a 6.3x speedup in our MIC-enabled code on a single compute node compared to baseline, which is quite near the maximum theoretical performance benefit. Additionally, a 64-compute-node run indicated a 5.6x speedup relative to baseline.





\balance

\bibliographystyle{abbrv}
\bibliography{part,ccgo}

\end{document}